\documentclass[11pt]{article}
\usepackage{amsmath, amsfonts,epsf,graphicx,times,appendix,multirow,hhline,color,float}
\usepackage{algorithm}
\usepackage{algorithmic}
\usepackage{subfigure}

\usepackage{amssymb}
\usepackage{amsthm}
\usepackage{mathrsfs}
\usepackage{extarrows}
\usepackage{xcolor}

\newcommand{\ls}[1]  %% 1 in brackets means \ls takes 1 argument
   {\dimen0=\fontdimen6\the=#1\dimen0
    \advance\lineskip.5\fontdimen5\the\lineskip-\dimen0
    \lineskiplimit=.9\lineskip
    \baselineskip=\lineskip
    \advance\baselineskip\dimen0
    \normallineskip\lineskip
    \normallineskiplimit\lineskiplimit
    \normalbaselineskip\baselineskip
    \ignorespaces
   }

\begin{document}

\title{\bf \LARGE Technical Report on Efficient Integration of Dynamic TDD with Massive MIMO}

\author{Yan Huang$^{\ast}$, Brian Jalaian$^{\dag}$, Stephen Russell$^{\dag}$, and Hooman Samani$^{\ddag}$
\smallskip\\
 \small The Bradley Department of Electrical and Computer Engineering, Virginia Tech, Blacksburg, VA, USA$^{\ast}$\\
 \small U.S. Army Research Laboratory, Adelphi, MD, USA$^\dag$\\
 \small The Department of Electrical Engineering, National Taipei University, Taipei, Taiwan$^\ddag$\\
}

\date{}
\maketitle

\ls{1.5}

\begin{abstract}
Recent advances in massive multiple-input multiple-output (MIMO) communication show that equipping base stations (BSs) with large arrays of antenna can significantly improve the performance of cellular networks.
Massive MIMO has the potential to mitigate the interference in the network and enhance the average throughput per user.
On the other hand, dynamic time division duplexing (TDD), which allows neighboring  cells to operate with different uplink (UL) and downlink (DL) sub-frame configurations, is a promising enhancement for the conventional static TDD. Compared with static TDD, dynamic TDD can offer more flexibility to accommodate various UL and DL traffic patterns  across different cells, but may result in additional interference among cells transmitting in different directions.
Based on the unique characteristics and properties of massive MIMO and dynamic TDD, we propose a marriage of these two techniques,
i.e., to have massive MIMO address the limitation of dynamic TDD in macro cell (MC) networks.
Specifically, we advocate that the benefits of dynamic TDD can be fully extracted in MC networks equipped with massive MIMO, i.e.,
the BS-to-BS interference can be effectively removed by increasing the number of BS antennas.
We provide detailed analysis using random matrix theory to show that the effect of the BS-to-BS interference
on uplink transmissions vanishes as the number of BS antennas per-user grows infinitely large.
Last but not least, we validate our analysis by numerical simulations.
\end{abstract}


%%%%%%%%%%%%%%%%%%%%%%%%%%%%%%%%%%%%%%%%%%%%%%%%%%%%%%%%%%%%
\section{Introduction}
\label{sec:intro}

%%%Introducing Masssive MIMO
Massive multiple-input multiple-output (MIMO) has attracted much attention from research community in recent years\cite{marker22}--\cite{marker11}.
Massive MIMO is typically considered as a communication system consisting of a base station (BS) equipped with a very large antenna array and a number of single-antenna or multiple-antenna user terminals (UTs).
The number of antennas at the BS is usually much larger than the number of UT antennas.
In practical scenarios, there might be hundreds of BS antennas serving tens of UTs within one cell simultaneously \cite{marker22}.
An appealing feature of massive MIMO is its ability of mitigating the interference in the system and enhancing the average per-user throughput enabled by the large antenna array at the BS.
In \cite{marker5}, Marzetta shows that the impacts of the uncorrelated noise and
the intra-cellular interference are vanishing with the increasing number of BS antennas per-UT.
However, the pilot contamination effect is still remaining. 
Moreover, previous studies show that the transmit power per-antenna at the BSs could be reduced significantly
in massive MIMO systems \cite{marker7,marker21,marker2}.
This feature is appealing to future communication networks where green issues would be a major concern \cite{marker33}.

On the other hand, time division duplexing (TDD) is widely adopted in cellular communication systems due to its flexibility and promising features.
A key feature of TDD is the ability to accommodate various patterns of uplink (UL) and downlink (DL) traffic in cellular networks
(which is not as easy in frequency division duplexing (FDD) system) \cite{marker15}.
There are two main TDD schemes:
(i) static TDD and (ii) dynamic TDD.
In static TDD, all the neighboring cells must transmit in the same direction (UL or DL) on each sub-frame according to a predefined schedule \cite{marker18}, \cite{marker19}.
While with dynamic TDD, each cell is allowed to configure UL and DL sub-frames adaptively based on its traffic condition, such that adjacent cells in a network are not necessarily operating in the same direction at a given time instant.
The latter approach is also referred to as ``cell-specific traffic adaptation'' and ``dynamic UL-DL configuration'' \cite{marker18}-\cite{marker4}.
Dynamic TDD can potentially achieve a higher spectrum efficiency (by intelligently scheduling UL and DL sub-frames for each cell based on its traffic condition) compared with static TDD, where a unique UL-DL configuration is employed across all cells in the network \cite{marker19}.

Prior work has shown that the DL performance of a cellular network can be improved by employing dynamic TDD \cite{marker3}.
The reason is that the DL signal-to-interference-plus-noise (SINR) of a cell would increase when neighboring cells change their transmissions from DL to UL, which results in a reduction of interference experienced by the considered cell (since transmit power of UTs is much lower than that of BSs).
The key challenge for implementing dynamic TDD in realistic networks is the interference management for UL transmissions. In a dynamic TDD network, when a BS is receiving UL transmissions from UTs, the DL transmissions from BSs in neighboring cells would strongly impact its UL SINR. Such interference is called \emph{BS-to-BS interference}, as it is from a BS transmitting in DL to another BS receiving in UL \cite{marker3}.
If the BS-to-BS interference cannot be properly mitigated, the resulted loss in UL performance may offset the potential benefits of dynamic TDD.
Previous works in \cite{marker3}, \cite{marker20}, and \cite{marker16} show that small cell (SC) deployments can actually benefit from employing dynamic TDD.
This is due to the fact that the BS-to-BS interference is negligible as the SC BSs typically
operate at lower transmit power compared to the macro cell (MC) BSs and are usually better mutually isolated.
However, employing dynamic TDD in MC is not recommended without appropriate BS-to-BS interference mitigation techniques.
The limitation of dynamic TDD in MC is attributed to higher transmit power of MC BSs and the line-of-sight (LoS) characteristic of channels between MC BSs (causing strong BS-to-BS couplings)\cite{marker3, marker1}.

Based on the unique characteristics and properties of massive MIMO and dynamic TDD, we propose a marriage of these two techniques,
i.e., to have massive MIMO address the limitation of dynamic TDD in MC networks.
Specifically, we advocate that the benefits of dynamic TDD can be fully extracted in MC networks equipped with massive MIMO, i.e.,
the BS-to-BS interference can be effectively removed when the number of BS antennas is very large.

The main contributions of this paper are as follows.
\begin{itemize}
 \item To the best of our knowledge, this is the first paper that considers employing massive MIMO to reap the benefits of dynamic TDD and address its limitations in MC networks. Previous works on massive MIMO mainly considered static TDD, which cannot accommodate uneven UL and DL traffic demands across different MCs. On the other hand, studies on dynamic TDD concluded that MC networks is not suitable for operating with dynamic TDD. In contrast, our analysis reveals the potential benefits from the marriage of both techniques to MC networks.
 \item By using the random matrix methods, we derive deterministic approximations of the BS-to-BS interference and per-user achievable rate in dynamic TDD networks. Based on the result, we show that the impact of the BS-to-BS interference on UL transmissions vanishes as the number of BS antennas per-UT grows infinitely large.
 \item We show that dynamic TDD in massive MIMO can increase the per-user average achievable rates in both UL and DL.
 \item We conduct numerical simulations to verify that a dynamic TDD network with massive MIMO achieves higher throughput in both DL and UL compared with a static TDD network.

\end{itemize}

The remainder of this paper is organized as follows.
In Section~\ref{sec:system_model}, we explain our system model in details.
In Section~\ref{sec:feasibility}, we derive a deterministic approximation of the power of BS-to-BS interference and show that this power decreases to zero as the number of BS antennas per UT increases infinitely.
In section~\ref{sec:numerical_resutl}, we validate our analysis by numerical simulations.
Section~\ref{sec:conclusion} concludes this paper.

%%%%%%%%%%%%%%%%%%%%%%%%%%%%%%%%%%%%%%%%%%%%%%%%%%%%%

\section{System Model}
\label{sec:system_model}
For notation, the operators $\mathop{\mathrm{tr}}(\cdot)$, $\mathop{\mathbb{E}}[ \cdot]$, $\mathop{\mathbb{E}}[ \cdot | \cdot]$, $(\cdot)^{\top}$ and $(\cdot)^{\dagger}$ represent trace, expectation, conditional expectation, transpose and Hermitian transpose, respectively. $\mathrm{lim}_{M}$ denotes $\mathrm{lim}_{M \to \infty}$. The notation $\mathcal{CN} (\mathbf{0},\mathbf{R})$ stands for the circular symmetric complex Gaussian distribution with mean $\mathbf{0}$ and covariance matrix $\mathbf{R}$.
The notation``$\xlongrightarrow[M\to \infty]{\mathrm{a.s.}}$" represents almost sure convergence as $M\to\infty$.

\begin{figure}
  \centering
  \includegraphics[width=0.5\textwidth]{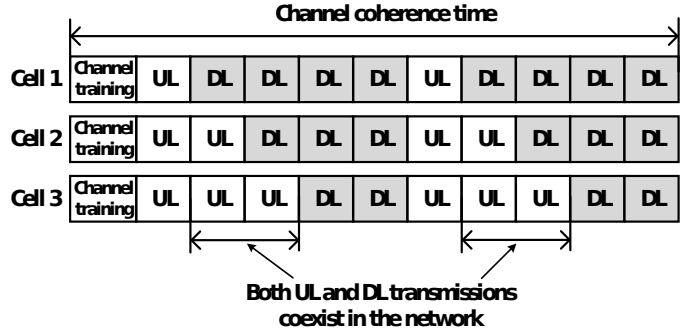}
  \caption{Dynamic cell-specific UL-DL configurations.} \label{fig:marker1}
\end{figure}

The cellular network consists of $L$ MCs with a frequency reuse factor of one. Each cell contains $K$ single-antenna UTs and an MC BS equipped with $M$ antennas. All channels are assumed to be flat-fading. In the time domain, we employ a block-fading channel model, i.e., channel fading is constant within the coherence time. Each individual coherence time interval is partitioned into two phases that are allocated for channel training and data transmission, respectively. The training and transmission phases of all $L$ cells are supposed to be perfectly synchronized, such that during a channel training phase, BSs would only receive training signals from UTs. During a transmission phase, each cell can schedule UL and DL sub-frames based on its own traffic condition (more UL sub-frames would be scheduled if UL traffic is heavier than DL traffic, and vice versa). Thus on a given sub-frame, some of the $L$ cells may be operating in UL while others are in DL. We exemplify such radio frame structure in Fig.\ref{fig:marker1}. The focus of our analysis in this paper is on sub-frames where transmissions of both UL and DL coexist in the network.
\subsection{Uplink Reception}
On a sub-frame when transmissions of both UL and DL coexist in the network, a BS operating in UL receives not only transmissions from UTs in UL cells, but also signals from BSs transmitting in DL. Denote the set of the indices of cells in UL by $\mathcal{S}_{u}$ and cells in DL by $\mathcal{S}_{d}$. Let $\mathbf{y}^{\mathrm{ul}}_{j} \in \mathbb{C}^{M}$ represents the instantaneous received base-band signal vector at BS $j \in \mathcal{S}_{u}$, then
\begin{equation}
\mathbf{y}^{\mathrm{ul}}_{j} = \sqrt{p_{\mathrm{ul}}} \sum_{l \in \mathcal{S}_{u}} \sum_{m=1}^{K} \mathbf{h}_{jlm} x_{lm} + \sqrt{p_{\mathrm{dl}}} \sum_{n \in \mathcal{S}_{d}} \mathbf{G}_{jn} \mathbf{W}_{n} \mathbf{z}_{n} + \mathbf{n}^{\mathrm{ul}}_{j},
\end{equation}
where $\mathbf{h}_{jlm} \in \mathbb{C}^{M}$ denotes the channel vector from UT $m$ in cell $l$ to BS $j$, $\mathbf{G}_{jn} \in \mathbb{C}^{M \times M}$ denotes the channel matrix from BS $n$ to BS $j$, $x_{lm} \sim \mathcal{CN} \left( 0,1 \right)$ is the independent transmit signal of UT $m$ in cell $l$, $\mathbf{z}_{n} \in \mathbb{C}^{K} \sim \mathcal{CN} \left( \mathbf{0}, \mathbf{I}_{K} \right)$ is the independent DL data symbol vector at BS $n$ for the $K$ UTs in cell $n$ \cite{marker30}, $\mathbf{n}^{\mathrm{ul}}_{j} \in \mathbb{C}^{M} \sim \mathcal{CN} \left( \mathbf{0}, \mathbf{I}_{M} \right)$ is the receiver noise vector that is uncorrelated with the channels and data signals, $\mathbf{W}_{n} \in \mathbb{C}^{M \times K}$ is the precoding matrix used for the DL transmissions at BS $n$, and $p_{\mathrm{ul}}$, $p_{\mathrm{dl}}$ models the UL and DL SNRs, respectively. The channel vector $\mathbf{h}_{jlm}$ is given as \cite{marker6}
\begin{equation} \label{eq:marker41}
\mathbf{h}_{jlm}=\bar{\mathbf{R}}_{jlm} \mathbf{v}_{jlm} \in \mathbb{C}^{M},
\end{equation}
where the deterministic Hermitian-symmetric positive definite matrix $\mathbf{R}_{jlm}=\bar{\mathbf{R}}_{jlm}\bar{\mathbf{R}}_{jlm}^{\dagger} \in \mathbb{C}^{M \times M}$ models the antenna correlation at the receiver and large-scale fading effects, and $\mathbf{v}_{jlm} \sim \mathcal{CN} \left( \mathbf{0},\mathbf{I}_{M} \right)$ is an independent Rayleigh fading channel vector \cite{marker31}. The channel matrix $\mathbf{G}_{jn}$ is modeled as \cite{marker8}
\begin{equation} \label{eq:marker9}
\mathbf{G}_{jn}=\breve{\mathbf{G}}_{jn}+\bar{\mathbf{G}}_{jn}=\bar{\mathbf{C}}_{jn} \mathbf{V}_{jn} \bar{\mathbf{T}}_{jn}+\bar{\mathbf{G}}_{jn} \in \mathbb{C}^{M \times M},
\end{equation}where $\breve{\mathbf{G}}_{jn}\in \mathbb{C}^{M \times M}$ and the deterministic matrix $\bar{\mathbf{G}}_{jn} \in \mathbb{C}^{M \times M}$ respectively correspond to the non-line-of-sight (NLoS) and the line-of-sight (LoS) components of the channel, $\mathbf{C}_{jn}=\bar{\mathbf{C}}_{jn}\bar{\mathbf{C}}_{jn}^{\dagger}\in \mathbb{C}^{M \times M}$ and $\mathbf{T}_{jn}=\bar{\mathbf{T}}_{jn}^{\dagger}\bar{\mathbf{T}}_{jn} \in \mathbb{C}^{M \times M}$ are deterministic matrices similar to $\mathbf{R}_{jlm}$ that characterize the large-scale fading and spatial correlation structures at receiving and transmitting antenna arrays, respectively, and $\mathbf{V}_{jn} \in \mathbb{C}^{M \times M}$ is a statistically independent random matrix whose entries are Gaussian and independent and identically distributed (i.i.d.) with zero mean and unit variance.
\subsection{Downlink Reception}
We assume perfect channel reciprocity in this paper, i.e., DL channels are simply Hermitian transposes of UL channels.

The UTs in a DL cell receive signals transmitted from all BSs operating in DL and UTs in UL cells. Let $\mathbf{y}^{\mathrm{dl}}_{ik} \in \mathbb{C}^{M}$ denote the instantaneous base-band receive signal at UT $k$ in cell $i$, where $i\in \mathcal{S}_{d }$. Then
\begin{equation}
y^{\mathrm{dl}}_{ik} = \sqrt{p_{\mathrm{dl}}} \sum_{n \in \mathcal{S}_{d}} \mathbf{h}^{\dagger}_{nik} \mathbf{W}_{n} \mathbf{z}_{n}+ \sqrt{p_{\mathrm{ul}}} \sum_{l \in \mathcal{S}_{u}} \sum_{m=1}^{K} g_{iklm} x_{lm} +n^{\mathrm{dl}}_{ik},
\end{equation}where $g_{iklm}\sim \mathcal{CN} \left( 0,\alpha_{iklm} \right)$ denotes the channel response from UT $m$ in cell $l$ to UT $k$ in cell $i$, $\alpha_{iklm}$ represents the transmit SNR and large-scale fading on $g_{iklm}$, and $n^{\mathrm{dl}}_{ik}\sim \mathcal{CN} \left( 0,1\right)$ is the receiver noise.
\subsection{Channel Estimation}
We consider uplink channel estimation in this paper. 
The channels $\mathbf{h}_{jlk}$'s are estimated by BSs through UL pilot signaling from UTs.
Under perfect channel reciprocity, DL channel estimates are simply Hermitian transposes of UL channel estimates.
In each coherence time interval, the channel training phases of all $L$ cells are assumed to be perfectly synchronized (see Fig.\ref{fig:marker1}). Thus during a channel training phase, the BSs would only receive training signals from UTs.

Assume that a set of $K$ mutually orthogonal pilot sequences is reused in all $L$ cells. 
By correlating the received pilot signal with the pilot sequence corresponding to UT $k$, BS $j$ obtains
\begin{equation}
\mathbf{y}^{\mathrm{tr}}_{jk} = \sum_{l \in \mathcal{S}_{u}}\mathbf{h}_{jlk}+\frac{1}{\sqrt{p_{\mathrm{tr}}}} \mathbf{n}_{jk}^{\mathrm{tr}},
\end{equation}where the noise vector $\mathbf{n}_{jk}^{\mathrm{tr}} \in \mathbb{C}^{M} \sim \mathcal{CN} \left( \mathbf{0}, \mathbf{I}_{M} \right)$ and $p_{\mathrm{tr}}$ denotes the SNR in the uplink training phase. From $\mathbf{y}^{\mathrm{tr}}_{jk}$, BS $j$ can further make an minimum mean square error (MMSE) estimate $\hat{\mathbf{h}}_{jjk}$ for the channel vector $\mathbf{h}_{jjk}$, which is given as \cite{marker6}
\begin{equation}
\hat{\mathbf{h}}_{jjk}=\mathbf{R}_{jjk}\mathbf{Q}_{jk}\left( \sum_{l=1}^{L} \mathbf{h}_{jlk} +\frac{1}{\sqrt{p_{\mathrm{tr}}}} \mathbf{n}_{jk}^{\mathrm{tr}} \right),
\end{equation}where $\mathbf{Q}_{jk}$ is defined as $\mathbf{Q}_{jk} = \left( \sum_{l=1}^{L} \mathbf{R}_{jlk} + \frac{1}{p_{\mathrm{tr}}} \mathbf{I}_{M} \right)^{-1}$. Denote $\mathbf{\Phi}_{jlmk}=\mathbf{R}_{jlk} \mathbf{Q}_{jk} \mathbf{R}_{jmk}$. It can be shown that $\hat{\mathbf{h}}_{jjk}$ is Gaussian distributed as $\hat{\mathbf{h}}_{jjk} \sim \mathcal{CN} \left( \mathbf{0}, \mathbf{\Phi}_{jjjk} \right)$. With the above MMSE estimate, the channel $\mathbf{h}_{jjk}$ can be decomposed as
\begin{equation} \label{eq:marker2}
\mathbf{h}_{jjk}= \hat{\mathbf{h}}_{jjk}+ \tilde{\mathbf{h}}_{jjk},
\end{equation}where $\tilde{\mathbf{h}}_{jjk} \sim \mathcal{CN} \left( \mathbf{0}, \mathbf{R}_{jjk}-\mathbf{\Phi}_{jjjk} \right)$ is the estimation error vector and uncorrelated with $\hat{\mathbf{h}}_{jjk}$. Furthermore, $\hat{\mathbf{h}}_{jjk}$ and $\tilde{\mathbf{h}}_{jjk}$ are statistically independent because the two vectors are jointly Gaussian distributed. 

The correlation matrices $\mathbf{R}_{jlk}$, $\mathbf{C}_{jl}$ and $\mathbf{T}_{jl}$ can be estimated by using standard covariance estimation techniques and therefore are supposed to be perfectly known by BSs.
The channel training between BSs for estimating the channel matrices $\mathbf{G}_{jl}$ is not considered in this work, which means that the instantaneous information of BS-to-BS channels are unknown to both ends.

\subsection{Uplink Detection and Downlink Precoding}
In this paper, we consider MMSE detection and precoding for UL and DL transmissions, respectively. Denote the MMSE detection vector corresponding to the UT $k$ in cell $j$ as \cite{marker6}
{\setlength\arraycolsep{1pt}
\begin{eqnarray} \label{eq:marker1}
\mathbf{a}_{jk}&=&\frac{1}{M}\mathbf{\Lambda}_{j}^{\mathrm{ul}}\hat{\mathbf{h}}_{jjk}\nonumber\\
&=&\frac{1}{M} \left( \frac{1}{M}\sum_{i=1}^{K} \hat{\mathbf{h}}_{jji} \hat{\mathbf{h}}_{jji}^{\dagger} +\frac{1}{M}\mathbf{F}_{j}^{\mathrm{ul}} + \varphi_{j}^{\mathrm{ul}} \mathbf{I}_{M} \right)^{-1}\hat{\mathbf{h}}_{jjk},\nonumber\\
\end{eqnarray}}where $\varphi_{j}^{\mathrm{ul}} > 0$ is a regularization parameter and $\mathbf{F}_{j}^{\mathrm{ul}} \in \mathbb{C}^{M \times M}$ is a Hermitian nonnegative definite matrix. The MMSE precoding matrix used for DL transmissions at BS $n$ is defined as
{\setlength\arraycolsep{1pt}
\begin{eqnarray} \label{eq:marker5}
\mathbf{W}_{n}&=&\sqrt{\lambda_{n}} \mathbf{\Omega}_{n}\nonumber\\
&=&\sqrt{\lambda_{n}}\left( \sum_{i=1}^{K} \hat{\mathbf{h}}_{nni} \hat{\mathbf{h}}_{nni}^{\dagger} +\mathbf{F}_{n}^{\mathrm{dl}} + M\varphi_{n}^{\mathrm{dl}} \mathbf{I}_{M} \right)^{-1} \hat{\mathbf{H}}_{nn},\nonumber\\
\end{eqnarray}}where $\hat{\mathbf{H}}_{nn} = \left[\hat{\mathbf{h}}_{nn1} \ldots  \hat{\mathbf{h}}_{nnK}\right]$, $\varphi_{n}^{\mathrm{dl}} > 0$ and $\mathbf{F}_{n}^{\mathrm{dl}} \in \mathbb{C}^{M \times M}$ are regularization parameters similar to those in (\ref{eq:marker1}), and $\lambda_{n}$ normalizes the expectation of the transmit power per UT of BS $n$ and is defined as
\begin{equation}
\lambda_{n}=\frac{K}{\mathop{\mathbb{E}} \left[ \mathop{\mathrm{tr}} \mathbf{\Omega}_{n} \mathbf{\Omega}_{n}^{\dagger} \right]}.
\end{equation}The parameters $\varphi_{j}^{\mathrm{ul}}$, $\mathbf{F}_{j}^{\mathrm{ul}}$, $\varphi_{n}^{\mathrm{dl}}$ and $\mathbf{F}_{n}^{\mathrm{dl}}$ could be optimized according to certain criterion, which is not addressed in this work. The setting for these quantities is arbitrary and has no impact on our analysis in Section~\ref{sec:feasibility}. For example, following standard approaches for deriving the MMSE detector and precoder, one could set $\varphi_{j}^{\mathrm{ul}}=\frac{1}{Mp_{\mathrm{ul}}}$, $\varphi_{n}^{\mathrm{dl}}=\frac{1}{Mp_{\mathrm{dl}}}$, and $\mathbf{F}_{j}^{\mathrm{ul}}$ and $\mathbf{F}_{n}^{\mathrm{dl}}$ as covariance matrices of the interference and error terms \cite{marker6}.

\subsection{Ergodic Achievable Rates}
In the UL transmission phase, BS $j$ processes its received signal $\mathbf{y}^{\mathrm{ul}}_{j}$ with the linear detection vector $\mathbf{a}_{jk}$ to obtain an estimate $\hat{x}_{jk}$ for the transmitted signal $x_{jk}$, i.e.,
{\setlength\arraycolsep{1pt}
\begin{eqnarray}
\hat{x}_{jk} & = & \mathbf{a}_{jk}^{\dagger} \mathbf{y}^{\mathrm{ul}}_{j} \nonumber\\
& = & \sqrt{p_{\mathrm{ul}}} \mathbf{a}_{jk}^{\dagger} \hat{\mathbf{h}}_{jjk} x_{jk}+\sqrt{p_{\mathrm{ul}}} \mathbf{a}_{jk}^{\dagger} \tilde{\mathbf{h}}_{jjk} x_{jk}\nonumber\\  & &+ \sqrt{p_{\mathrm{ul}}} \sum_{l \in \mathcal{S}_{u}} \sum_{\scriptstyle m=1 \atop \scriptstyle m\neq k}^{K} \mathbf{a}_{jk}^{\dagger} \mathbf{h}_{jlm} x_{lm} \nonumber\\
& &+ \sqrt{p_{\mathrm{ul}}} \sum_{\scriptstyle l \in \mathcal{S}_{u} \atop \scriptstyle l \neq j} \mathbf{a}_{jk}^{\dagger} \mathbf{h}_{jlk} x_{lk} + \sqrt{p_{\mathrm{dl}}} \sum_{n \in \mathcal{S}_{d}} \mathbf{a}_{jk}^{\dagger} \breve{\mathbf{G}}_{jn} \mathbf{W}_{n} \mathbf{z}_{n} \nonumber\\
& & +\sqrt{p_{\mathrm{dl}}} \sum_{n \in \mathcal{S}_{d}} \mathbf{a}_{jk}^{\dagger} \bar{\mathbf{G}}_{jn} \mathbf{W}_{n} \mathbf{z}_{n}+ \mathbf{a}_{jk}^{\dagger}\mathbf{n}_{j},
\end{eqnarray}}where the decompositions of $\mathbf{G}_{jn}$ and $\mathbf{h}_{jjk}$ follow from (\ref{eq:marker9}) and (\ref{eq:marker2}), respectively. The associated UL SINR $\gamma^{\mathrm{ul}}_{jk}$ takes the form
\begin{equation}\label{eq:UL_SINR}
\gamma^{\mathrm{ul}}_{jk}=\frac{S^{\mathrm{ul}}_{jk}}{I_{jk}^{\mathrm{ul},(1)}+I_{jk}^{\mathrm{ul},(2)}+I_{jk}^{\mathrm{ul},(3)}+I_{jk}^{\mathrm{ul},(4)}+I_{jk}^{\mathrm{ul},(5)}},
\end{equation}
where
{\setlength\arraycolsep{1pt}
\begin{eqnarray}
S^{\mathrm{ul}}_{jk} & = &p_{\mathrm{ul}} \lvert \mathbf{a}_{jk}^{\dagger} \hat{\mathbf{h}}_{jjk} \rvert ^{2}, \\
I_{jk}^{\mathrm{ul},(1)} & = &p_{\mathrm{ul}} \mathop{\mathbb{E}} \left[ \mathbf{a}_{jk}^{\dagger} \tilde{\mathbf{h}}_{jjk} \tilde{\mathbf{h}}_{jjk}^{\dagger} \mathbf{a}_{jk} \bigg| \hat{\mathbf{H}}_{jj} \right], \\
I_{jk}^{\mathrm{ul},(2)} & = &p_{\mathrm{ul}} \sum_{l \in \mathcal{S}_{u}} \sum_{\scriptstyle m=1 \atop \scriptstyle m\neq k}^{K} \mathop{\mathbb{E}} \left[ \mathbf{a}_{jk}^{\dagger} \mathbf{h}_{jlm} \mathbf{h}_{jlm}^{\dagger} \mathbf{a}_{jk} \bigg| \hat{\mathbf{H}}_{jj} \right],\\
I_{jk}^{\mathrm{ul},(3)}&=& p_{\mathrm{ul}}\sum_{\scriptstyle l \in \mathcal{S}_{u} \atop \scriptstyle l \neq j} \mathop{\mathbb{E}} \left[ \mathbf{a}_{jk}^{\dagger} \mathbf{h}_{jlk} \mathbf{h}_{jlk}^{\dagger} \mathbf{a}_{jk} \bigg| \hat{\mathbf{H}}_{jj} \right], \\
I_{jk}^{\mathrm{ul},(4)} & = & p_{\mathrm{dl}} \sum_{n \in \mathcal{S}_{d}} \Bigg\{ \mathop{\mathbb{E}} \left[ \mathbf{a}_{jk}^{\dagger} \breve{\mathbf{G}}_{jn} \mathbf{W}_{n} \mathbf{W}_{n}^{\dagger} \breve{\mathbf{G}}_{jn}^{\dagger} \mathbf{a}_{jk} \bigg| \hat{\mathbf{H}}_{jj} \right] \label{eq:marker53}\nonumber\\
&&+ \mathop{\mathbb{E}} \left[ \mathbf{a}_{jk}^{\dagger} \bar{\mathbf{G}}_{jn} \mathbf{W}_{n} \mathbf{W}_{n}^{\dagger} \bar{\mathbf{G}}_{jn}^{\dagger} \mathbf{a}_{jk} \bigg| \hat{\mathbf{H}}_{jj} \right]\Bigg\},\\
I_{jk}^{\mathrm{ul},(5)} & = & \mathbf{a}_{jk}^{\dagger} \mathbf{a}_{jk}. \label{eq:marker19}
\end{eqnarray}}The terms $I_{jk}^{\mathrm{ul},(1)}$, $I_{jk}^{\mathrm{ul},(2)}$, $I_{jk}^{\mathrm{ul},(3)}$, $I_{jk}^{\mathrm{ul},(4)}$ and $I_{jk}^{\mathrm{ul},(5)}$ stand for different categories of interference and noise. 
Specifically, $I_{jk}^{\mathrm{ul},(1)}$ corresponds to the channel estimate error. $I_{jk}^{\mathrm{ul},(2)}$ and $I_{jk}^{\mathrm{ul},(3)}$ are the inter-UT interference from all other UTs transmitting in UL in the network. $I_{jk}^{\mathrm{ul},(5)}$ is the receiver noise. $I_{jk}^{\mathrm{ul},(4)}$ characterizes the BS-to-BS interference caused by DL transmissions from neighboring BSs.
$I_{jk}^{\mathrm{ul},(4)}$ is unique to dynamic TDD and will disappear when static TDD is employed.

The UL ergodic achievable rate $R^{\mathrm{ul}}_{jk}$ of UT $k$ in cell $j$ is defined as \cite{marker6}, \cite{marker9}
\begin{equation} \label{eq:marker3}
R^{\mathrm{ul}}_{jk}=\mathop{\mathbb{E}} \left[ \mathrm{log}_{2} \left( 1+ \gamma^{\mathrm{ul}}_{jk} \right) \right].
\end{equation}

As for the DL, we decompose the DL received signal $y^{\mathrm{dl}}_{ik}$ as
{\setlength\arraycolsep{1pt}
\begin{eqnarray}
y^{\mathrm{dl}}_{ik}& =&\sqrt{p_{\mathrm{dl}}}\mathop{\mathbb{E}} \left[\mathbf{h}^{\dagger}_{iik} \mathbf{w}_{ik} \right]z_{ik}\nonumber\\
&&+\sqrt{p_{\mathrm{dl}}}\left( \mathbf{h}^{\dagger}_{iik} \mathbf{w}_{ik}-\mathop{\mathbb{E}} \left[\mathbf{h}^{\dagger}_{iik} \mathbf{w}_{ik} \right] \right)z_{ik} \nonumber\\
&&+\sqrt{p_{\mathrm{dl}}}\sum_{\scriptstyle m=1 \atop \scriptstyle m\neq k}^{K}\mathbf{h}^{\dagger}_{iik} \mathbf{w}_{im}z_{im}+ \sqrt{p_{\mathrm{dl}}} \sum_{\scriptstyle n \in \mathcal{S}_{d} \atop \scriptstyle n\neq i } \mathbf{h}^{\dagger}_{nik} \mathbf{W}_{n} \mathbf{z}_{n}\nonumber\\
&&+ \sqrt{p_{\mathrm{ul}}} \sum_{l \in \mathcal{S}_{u}} \sum_{m=1}^{K} g_{iklm} x_{lm} +n^{\mathrm{dl}}_{ik},
\end{eqnarray}}where $\mathbf{w}_{ik}$ is the $k$th column of $\mathbf{W}_{i}$ and $z_{ik}$ is the $k$th entry of $\mathbf{z}_{i}$. By treating $\mathop{\mathbb{E}} \left[\mathbf{h}^{\dagger}_{iik} \mathbf{w}_{ik} \right]$ as the average effective channel and assuming that it is perfectly known at the UT, we can write the DL ergodic achievable rate as
\begin{equation}
R^{\mathrm{dl}}_{jk}=\mathrm{log}_{2} \left( 1+ \gamma^{\mathrm{dl}}_{jk} \right),
\end{equation}where the DL SINR
\begin{equation}
\gamma^{\mathrm{dl}}_{jk}=\frac{S^{\mathrm{dl}}_{jk}}{I_{jk}^{\mathrm{dl},(1)}+I_{jk}^{\mathrm{dl},(2)}+I_{jk}^{\mathrm{dl},(3)}+I_{jk}^{\mathrm{dl},(4)}+1},
\end{equation}where
{\setlength\arraycolsep{1pt}
\begin{eqnarray}
S^{\mathrm{dl}}_{jk} & = &p_{\mathrm{dl}} \Big\lvert \mathop{\mathbb{E}} \left[\mathbf{h}^{\dagger}_{iik} \mathbf{w}_{ik} \right] \Big\rvert ^{2}, \\
I_{jk}^{\mathrm{dl},(1)} & = &p_{\mathrm{dl}} \mathop{\mathrm{var}} \left[ \mathbf{h}^{\dagger}_{iik} \mathbf{w}_{ik}  \right], \\
I_{jk}^{\mathrm{dl},(2)} & = &p_{\mathrm{dl}} \sum_{\scriptstyle m=1 \atop \scriptstyle m\neq k}^{K}\mathop{\mathbb{E}} \left[ \mathbf{h}^{\dagger}_{iik} \mathbf{w}_{im} \mathbf{w}^{\dagger}_{im}\mathbf{h}_{iik} \right], \\
I_{jk}^{\mathrm{dl},(3)}&=& p_{\mathrm{dl}}\sum_{\scriptstyle n \in \mathcal{S}_{d} \atop \scriptstyle n\neq i }\mathop{\mathbb{E}} \left[ \mathbf{h}^{\dagger}_{nik} \mathbf{W}_{n} \mathbf{W}^{\dagger}_{n}\mathbf{h}_{nik} \right], \\
I_{jk}^{\mathrm{dl},(4)} & = & p_{\mathrm{ul}} \sum_{l \in \mathcal{S}_{u}} \sum_{m=1}^{K} \alpha_{iklm}. \label{eq:U2Uinterf}
\end{eqnarray}}
$I_{jk}^{\mathrm{dl},(4)}$ represents the UT-to-UT interference from all UTs transmitting in UL in the network.
\section{The Feasibility of Dynamic TDD in Massive MIMO Networks}
\label{sec:feasibility}
In this section, we study the feasibility of dynamic TDD in massive MIMO MC networks by taking a closer look at the interference experienced by each UL or DL link.

In fact, there would be a trade-off among different categories of interference effects when employing dynamic TDD in a cellular network.
That is, the interference in one category increases while it decreases in the other. 
To illustrate this trade- off, consider a cell operating in UL.
With dynamic TDD, the BS-to-BS interference $I_{jk}^{\mathrm{ul},(4)}$ from adjacent DL cells increases while the interference $I_{jk}^{\mathrm{ul},(2)}$, $I_{jk}^{\mathrm{ul},(3)}$ decreases. 
Similarly, for a cell operating in DL with dynamic TDD, the UT-to-UT interference $I_{jk}^{\mathrm{dl},(4)}$ from adjacent UL cells increases while the interference $I_{jk}^{\mathrm{dl},(3)}$ decreases.

Practically, the transmit power of UTs is much lower than that of BSs. In addition, the channels between UTs from different cells are more likely to be NLoS compared with inter-cell BS-to-UT channels. Thus when using dynamic TDD, the increase of $I_{jk}^{\mathrm{dl},(4)}$ would typically be less than the reduction of $I_{jk}^{\mathrm{dl},(3)}$, which means that the DL SINR (also the DL achievable rate) of dynamic TDD systems could be improved over that of static TDD systems.
This will be verified by the simulation results in Section \ref{sec:numerical_resutl}.

In UL, however, because of the large BS transmit power and the high probability that channels between BSs have LoS components, the interference $I_{jk}^{\mathrm{ul},(4)}$ might be much more significant than the reduction of $I_{jk}^{\mathrm{ul},(2)}$ and $I_{jk}^{\mathrm{ul},(3)}$. That is, the UL SINR might deteriorate with the application of dynamic TDD.

In order to investigate the UL performance of massive MIMO networks with dynamic TDD, we present an asymptotic analysis of the BS-to-BS interference.
The following assumptions are made:

\emph{Assumption 1:} The antenna correlation matrices $\mathbf{R}_{jlk}$, $\mathbf{C}_{jl}$ and $\mathbf{T}_{jl}$ are nonnegative definite and have uniformly bounded spectral norms on $M$, i.e., $\lim \mathrm{sup}_{M} \;\Vert \mathbf{R}_{jlk} \Vert \leq R < \infty \; \forall j,l,k$, $\lim \mathrm{sup}_{M} \;\Vert \mathbf{C}_{jl} \Vert \leq C < \infty \; \forall j,l$, and $\lim \mathrm{sup}_{M} \;\Vert \mathbf{T}_{jl} \Vert \leq T < \infty \; \forall j,l$.

\emph{Assumption 2:} The traces of the antenna correlation matrices $\mathbf{R}_{jlk}$, $\mathbf{C}_{jl}$ and $\mathbf{T}_{jl}$ are scaled up with respect to $M$, such that $\lim \mathrm{inf}_{M} \;\frac{1}{M} \mathop{\mathrm{tr}} \mathbf{R}_{jlk}\geq \epsilon_{R} > 0 \; \forall j,l,k$, $\lim \mathrm{inf}_{M} \;\frac{1}{M} \mathop{\mathrm{tr}} \mathbf{C}_{jl}\geq \epsilon_{C} > 0 \; \forall j,l$ and $\lim \mathrm{inf}_{M} \;\frac{1}{M} \mathop{\mathrm{tr}} \mathbf{T}_{jl}\geq \epsilon_{T} > 0 \; \forall j,l$.

\emph{Assumption 3:} The product LoS channel matrices $\frac{1}{M}\bar{\mathbf{G}}_{jn}^{\dagger}\bar{\mathbf{G}}_{jn}$ have uniformly bounded spectral norms on $M$, i.e., $\lim \mathrm{sup}_{M}\; \left\Vert \frac{1}{M} \bar{\mathbf{G}}_{jn}^{\dagger}\bar{\mathbf{G}}_{jn} \right\Vert \leq G < \infty \; \forall j\neq n$.

\emph{Assumption 4:} The parameter matrices $\mathbf{F}_{j}^{\mathrm{ul}}$ and $\mathbf{F}_{n}^{\mathrm{dl}}$ of the MMSE detector and the RZF precoder are Hermitian nonnegative definite and have uniformly bounded spectral norms on $M$, i.e., $\lim \mathrm{sup}_{M}\; \Vert \mathbf{F}_{j}^{\mathrm{ul}} \Vert < \infty \; \forall j$ and $\lim \mathrm{sup}_{M}\; \Vert \mathbf{F}_{n}^{\mathrm{dl}} \Vert < \infty \; \forall n$.

These assumptions are general in studies based on random matrix methods \cite{marker6,marker8,marker9,marker11}. Assumption 1 implies the conservation of channel energy \cite{marker11}, and enables the application of random matrix methods. The physical meaning of Assumption 2 is that the ranks of correlation matrices,  which represent degrees of freedom (DoF) of channels, increase at least proportionally with respect to $M$. Assumption 3 is necessary for the asymptotic analysis of the BS-to-BS interference, and holds true when the channel model for LoS channel components and design criteria of antenna arrays proposed in \cite{marker12}, \cite{marker13} and \cite{marker14} are applied. Under Assumption 3, we model the LoS matrices $\bar{\mathbf{G}}_{jn}$ as
\begin{equation}\label{eq:marker70}
\left[\bar{\mathbf{G}}_{jn}\right]_{r,c}=\alpha^{1/2}e^{\mathbf{i}\phi_{rc}}\;\;\;r,c\in\{1,2,\ldots,M\},
\end{equation}where $\alpha$ represents large-scale fading, and $\phi_{rc}$ is the phase of the $(r,c)$th entry. It has been shown that in certain conditions the columns of $\bar{\mathbf{G}}_{jn}$ can be orthogonal, such that $\bar{\mathbf{G}}_{jn}^{\dagger}\bar{\mathbf{G}}_{jn}=\alpha M\mathbf{I}_{M}$ and $\left\Vert \frac{1}{M} \bar{\mathbf{G}}_{jn}^{\dagger}\bar{\mathbf{G}}_{jn} \right\Vert=\alpha< \infty$. In this case, $\bar{\mathbf{G}}_{jn}$'s are full-rank $M\times M$ matrices. Assumption 4 holds since the matrices $\mathbf{F}_{j}^{\mathrm{ul}}$ and $\mathbf{F}_{n}^{\mathrm{dl}}$ are typically linear combinations of $\mathbf{R}_{jlk}$, $\mathbf{T}_{jl}$, $\mathbf{C}_{jl}$ and $\mathbf{\Phi}_{jjjk}$ (see, e.g., Eq. (12) in \cite{marker6}) whose spectral norms are uniformly bounded on $M$ under Assumption 1.
\subsection{Deterministic Approximation of the BS-to-BS interference}
In the following proposition, we provide a deterministic approximation $\bar{I}_{jk}^{\mathrm{ul},(4)}$ of the BS-to-BS interference $I_{jk}^{\mathrm{ul},(4)}$ based on random matrix methods Lemma 1 and Lemma 2 (see Appendix \ref{appendix_B}).

\emph{Proposition 1:} Let Assumptions 1-4 hold and $M \to \infty$ while $K/M < \infty$. Then, $I_{jk}^{\mathrm{ul},(4)}-\bar{I}_{jk}^{\mathrm{ul},(4)}\xlongrightarrow[M\to \infty]{\mathrm{a.s.}}0$, where $\bar{I}_{jk}^{\mathrm{ul},(4)}$ is given as
{\setlength\arraycolsep{1pt}
\begin{eqnarray}
&&\bar{I}_{jk}^{\mathrm{ul},(4)} \nonumber\\
&=&\frac{p_{\mathrm{dl}}}{\left( 1+ \frac{1}{M} \mathop{\mathrm{tr}} \mathbf{\Phi}_{jjjk}  \mathbf{\Psi}_{j} \right)^{2}} \times\nonumber\\
&&\sum_{n \in \mathcal{S}_{d}}\frac{\bar{\lambda}_{n}}{M} \Bigg\{\frac{1}{M}\sum_{m=1}^{K} \frac{\frac{1}{M} \mathop{\mathrm{tr}} \bar{\mathbf{G}}_{jn}^{\dagger} \bar{\mathbf{\Psi}}_{jk}^{\prime} \bar{\mathbf{G}}_{jn}\bar{\mathbf{\Gamma}}_{nm}^{\prime} }{\left( 1 + \frac{1}{M} \mathop{\mathrm{tr}} \mathbf{\Phi}_{nnnm} \mathbf{\Gamma}_{n} \right)^{2}}\nonumber\\ &&+ \left[ \sum_{m=1}^{K} \frac{\frac{1}{M} \mathop{\mathrm{tr}} \mathbf{\Phi}_{nnnm} \mathbf{\Gamma}_{jn}^{\prime}}{\left( 1 + \frac{1}{M} \mathop{\mathrm{tr}} \mathbf{\Phi}_{nnnm} \mathbf{\Gamma}_{n} \right)^{2}} \right] \frac{1}{M} \mathop{\mathrm{tr}} \mathbf{\Phi}_{jjjk} \mathbf{\Psi}_{jn}^{\prime} \Bigg\}, \nonumber\\
\label{eq:marker40}
\end{eqnarray}}where
\begin{equation}
\bar{\lambda}_{n} = \frac{K}{\sum_{m=1}^{K} \frac{\frac{1}{M} \mathop{\mathrm{tr}} \mathbf{\Phi}_{nnnm} \bar{\mathbf{\Gamma}}_{n}^{\prime}}{\left( 1 + \frac{1}{M} \mathop{\mathrm{tr}} \mathbf{\Phi}_{nnnm} \mathbf{\Gamma}_{n} \right)^{2}}},
\end{equation}and
\begin{enumerate}
\item $\mathbf{\Gamma}_{n} = \mathbf{\Gamma} \left( \varphi_{n}^{\mathrm{dl}} \right)$ is given by Lemma 1 for $\mathbf{R}_{i}=\mathbf{\Phi}_{nnni}\;\forall i$ and $\mathbf{S}=\mathbf{F}_{n}^{\mathrm{dl}}/M$,
\item $\mathbf{\Gamma}_{jn}^{\prime} = \mathbf{\Gamma}^{\prime} \left( \varphi_{n}^{\mathrm{dl}} \right)$ is given by Lemma 2 for $\mathbf{R}_{i}=\mathbf{\Phi}_{nnni}\;\forall i$, $\mathbf{S}=\mathbf{F}_{n}^{\mathrm{dl}}/M$ and $\mathbf{\Theta}=\mathbf{T}_{jn}$,
\item $\bar{\mathbf{\Gamma}}_{n}^{\prime} = \mathbf{\Gamma}^{\prime} \left( \varphi_{n}^{\mathrm{dl}} \right)$ is given by Lemma 2 for $\mathbf{R}_{i}=\mathbf{\Phi}_{nnni}\;\forall i$, $\mathbf{S}=\mathbf{F}_{n}^{\mathrm{dl}}/M$ and $\mathbf{\Theta}=\mathbf{I}_{M}$,
\item $\bar{\mathbf{\Gamma}}_{nm}^{\prime} = \mathbf{\Gamma}^{\prime} \left( \varphi_{n}^{\mathrm{dl}} \right)$ is given by Lemma 2 for $\mathbf{R}_{i}=\mathbf{\Phi}_{nnni}\;\forall i$, $\mathbf{S}=\mathbf{F}_{n}^{\mathrm{dl}}/M$ and $\mathbf{\Theta}=\mathbf{\Phi}_{nnnm}$,
\item $\mathbf{\Psi}_{j} = \mathbf{\Gamma} \left( \varphi_{j}^{\mathrm{ul}} \right)$ is given by Lemma 1 for $\mathbf{R}_{i}=\mathbf{\Phi}_{jjji}\;\forall i$ and $\mathbf{S}=\mathbf{F}_{j}^{\mathrm{ul}}/M$,
\item $\mathbf{\Psi}_{jn}^{\prime} = \mathbf{\Gamma}^{\prime} \left( \varphi_{j}^{\mathrm{ul}} \right)$ is given by Lemma 2 for $\mathbf{R}_{i}=\mathbf{\Phi}_{jjji}\;\forall i$, $\mathbf{S}=\mathbf{F}_{j}^{\mathrm{ul}}/M$ and $\mathbf{\Theta}=\mathbf{C}_{jn}$,
\item $\bar{\mathbf{\Psi}}_{jk}^{\prime} = \mathbf{\Gamma}^{\prime} \left( \varphi_{j}^{\mathrm{ul}} \right)$ is given by Lemma 2 for $\mathbf{R}_{i}=\mathbf{\Phi}_{jjji}\;\forall i$, $\mathbf{S}=\mathbf{F}_{j}^{\mathrm{ul}}/M$ and $\mathbf{\Theta}=\mathbf{\Phi}_{jjjk}$.
\end{enumerate}

Following similar approaches, we can derive deterministic approximations for all other terms in (\ref{eq:UL_SINR}), which are omitted in this paper to save space since our focus is on the BS-to-BS interference. Readers who are interested in those results may refer to \cite{marker6}. By replacing all terms in (\ref{eq:UL_SINR}) with their deterministic approximations, we obtain an approximation of the UL SINR denoted by $\bar{\gamma}^{\mathrm{ul}}_{jk}$. Then the UL ergodic achievable rate $R^{\mathrm{ul}}_{jk}$ can be approximated by $\bar{R}^{\mathrm{ul}}_{jk}=\mathrm{log}_{2} \left( 1+ \bar{\gamma}^{\mathrm{ul}}_{jk} \right)$. In Section \ref{sec:numerical_resutl}, we will use numerical results to verify the accuracy of $\bar{I}_{jk}^{\mathrm{ul},(4)}$ and $\bar{R}^{\mathrm{ul}}_{jk}$.

\subsection{Potential Improvement of UL Performance}

The channel model considered in the above analysis is general in the sense that various propagation characteristics and per-channel spatial correlations can be encompassed. However, it is difficult to physically understand and to gain insights from (\ref{eq:marker40}) under such channel model. Next, we consider a simplified case where all channels are assumed to be uncorrelated. Specifically, let
$\bar{\mathbf{R}}_{jjk}=\mathbf{I}_{M}, \bar{\mathbf{R}}_{jlk}=\sqrt{\alpha}\mathbf{I}_{M}, \bar{\mathbf{C}}_{jl}=\sqrt{\alpha}\mathbf{I}_{M}, \bar{\mathbf{T}}_{jl}=\mathbf{I}_{M}$ for $\forall l \neq j$, where $\alpha \in (0,1]$ represents the relative strength of the inter-cell interference. Additionally, the LoS matrices $\bar{\mathbf{G}}_{jl}$'s are modeled as (\ref{eq:marker70}), which satisfies $\frac{1}{M} \mathop{\mathrm{tr}} \bar{\mathbf{G}}_{jl}^{\dagger}\bar{\mathbf{G}}_{jl}=\alpha M$. With these assumptions, results of Lemma 1 and Lemma 2 can be given in closed form \cite{marker9}, and we have the following proposition:

\emph{Proposition 2:} Under the simplified channel model $\mathbf{F}_{n}^{\mathrm{dl}}=\mathbf{0}\;\forall n$, $\mathbf{F}_{j}^{\mathrm{ul}}=\mathbf{0}\;\forall j$ and $\varphi_{j}^{\mathrm{ul}}=\varphi>0\;\forall j$, $\bar{I}_{jk}^{\mathrm{ul},(4)}$ is given as
\begin{equation}\label{eq:marker24}
\bar{I}_{jk}^{\mathrm{ul},(4)}=2 p_{\mathrm{dl}} L_{\mathrm{dl}}\frac{\alpha\eta \tau^{\prime}K}{(1+\tau)^{2}M},
\end{equation}where $L_{\mathrm{dl}}$ is the number of cells operating in DL, and
$\eta=1+\alpha (L-1)+\frac{1}{p_{\mathrm{tr}}}, \tau=\frac{1-\varphi\eta -\kappa+\sqrt{\varphi^{2}\eta^{2}+(\kappa-1)^{2}+2\varphi \eta(\kappa+1)}}{2\varphi \eta}, \tau^{\prime}=\frac{(1+\tau)^{2}}{\kappa^{2}-\kappa+\varphi^{2}\eta^{2}(1+\tau)^{2}+2\varphi\eta\kappa(1+\tau)}$, with $\kappa=\frac{K}{M}$.

We can see from (\ref{eq:marker24}) that the BS-to-BS interference $\bar{I}_{jk}^{\mathrm{ul},(4)}$ is a function of the ratio $K/M$. 
If $K$ and $M$ increase with a fixed ratio $c\in(0,1)$, i.e., the number of BS antennas per-UT is a constant $c$, the BS-to-BS interference does not vanish as $M$ grows, i.e.,
$\bar{I}_{jk}^{\mathrm{ul},(4)}\to2 p_{\mathrm{dl}} L_{\mathrm{dl}}c\frac{\alpha\eta \bar{\tau}^{\prime}}{(1+\bar{\tau})^{2}}$, where $\bar{\tau}$ and $\bar{\tau}^{\prime}$ are obtained by substituting $c$ for $\kappa$ in $\tau$ and $\tau^{\prime}$. However, if $K$ is fixed or increases much more slowly than $M$ such that $K/M \to 0$, we have
$\bar{I}_{jk}^{\mathrm{ul},(4)}\to 0$,
i.e., the BS-to-BS interference can be removed by increasing $M$ to infinity.

In fact, the vanishing effect of the BS-to-BS interference also holds under the general channel model.

\emph{Proposition 3:} Under the general channel model, if there exists some $\delta>0$ such that $\lim \mathrm{inf}_{M}\;\varphi_{j}^{\mathrm{ul}}\geqslant\delta>0$ for all $j\in\mathcal{S}_{u}$, then both NLoS and LoS components of the BS-to-BS interference $I_{jk}^{\mathrm{ul},(4)}$ converge to zero almost surely as $M\to\infty$ while $K$ keeps constant.

\begin{proof}
The proof of Proposition 3 is given in Appendix \ref{appendix_A}.
\end{proof}

Previous work on massive MIMO systems has shown that as the number of antennas grows to infinity, all effects of channel estimation error $I_{jk}^{\mathrm{ul},(1)}$, interference $I_{jk}^{\mathrm{ul},(2)}$ and noise $I_{jk}^{\mathrm{ul},(5)}$ vanish except for the inter-cell interference $I_{jk}^{\mathrm{ul},(3)}$ due to pilot contamination \cite{marker5}, \cite{marker6}. 
This means that the interference from a UL cell cannot be entirely removed by increasing the number of BS antennas. However, our analysis shows that the BS-to-BS interference from a DL cell would vanish completely when $K/M\to 0$. Thus for a massive MIMO network, dynamic TDD can potentially improve achievable rates of UL cells as their adjacent interfering cells may change transmission directions from UL to DL and their received interference would decrease.

\section{Numerical Results}
\label{sec:numerical_resutl}
In this section, we validate our analysis in Section~\ref{sec:feasibility} by simulation results.
The network setting is as follows.
An inter-cell interference coefficient $\alpha=0.1$ is used to model the large-scale fading between neighboring cells.
The large-scale fading between a BS and UTs in its cell is normalized to one.
We employ the exponential model \cite{marker17, marker32} for antenna correlation matrices.
The correlation between adjacent antennas is denoted by $\beta$.
We consider $L=7$ and $K=10$.
Other parameters are set as $p_{\mathrm{tr}}=p_{\mathrm{ul}}=p_{\mathrm{dl}}=6\;\mathrm{dB}$, $\varphi_{j}^{\mathrm{ul}}=1/p_{\mathrm{ul}}$, $\varphi_{n}^{\mathrm{dl}}=1/p_{\mathrm{dl}}$, and $\mathbf{F}_{j}^{\mathrm{ul}}=\mathbf{F}_{j}^{\mathrm{dl}}=\mathbf{0}$.
%%%%%%%%%%%%%%%%%%%%%%%%%%%%%%%%%%%%%%%%%%%%%%%%%%%%%%%%%%%%%%
\begin{figure}[!t]
 \centering
  \includegraphics[width=0.33 \textwidth]{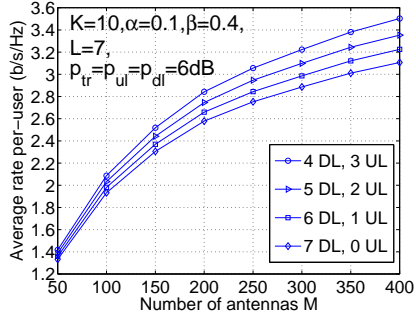}
 \caption[]{DL per-user achievable rates under different UL/DL cell assignments with antenna correlation $\beta=0.4$.}
\label{fig:DL}
\end{figure}

\begin{figure*}[!t]
 \centering
      \subfigure[$\beta=0.4$]{
   \includegraphics[width=0.33 \textwidth]{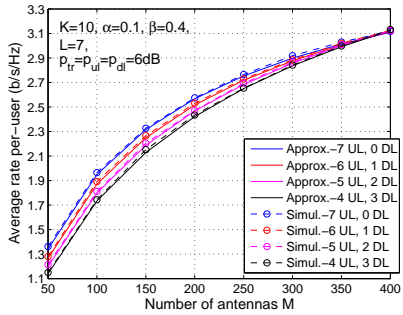}
    \label{fig:RateUL}
    }
    \hfill \hspace{-2em}
  \subfigure[$\beta=0.2$]{
  \includegraphics[width=0.33 \textwidth]{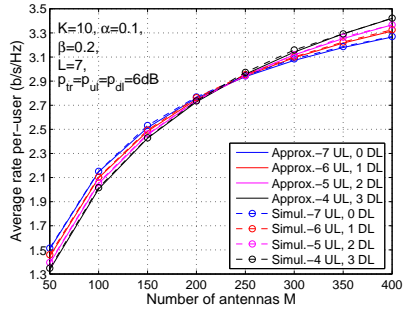}
   \label{fig:RateDLbeta2}
   }
    \hfill \hspace{-2em}
    \subfigure[$\beta=0$ (uncorrelated) ]{
   \includegraphics[width=0.33 \textwidth]{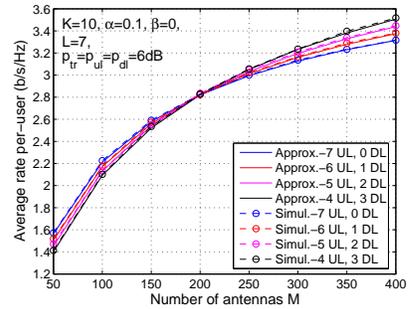}
    \label{fig:RateDLbeta0}
    }
 \caption[]{ Comparison of UL per-user achievable rates under different UL/DL cell assignments and antenna correlations. 
}
\label{fig:RateULdiffBeta}
\end{figure*}

\begin{figure*}[!t]
 \centering
 \subfigure[$\beta=0.4$]{
  \includegraphics[width=0.33 \textwidth]{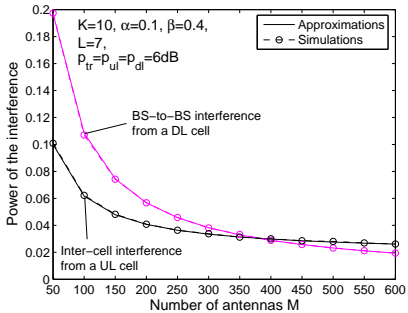}
   \label{fig:IntPwrbeta4}
   }
    \hfill \hspace{-2em}
     \subfigure[$\beta=0.2$]{
   \includegraphics[width=0.33 \textwidth]{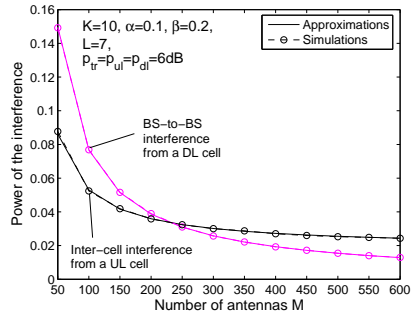}
    \label{fig:IntPwrbeta2}
    }
   \hfill \hspace{-2em}
    \subfigure[$\beta=0$ (uncorrelated)]{
  \includegraphics[width=0.33 \textwidth]{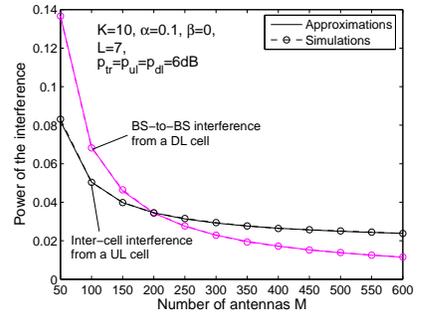}
   \label{fig:IntPwrbeta0}
   }
 \caption[]{ Comparison between the BS-to-BS interference from a DL cell and the inter-cell interference from a UL cell under different antenna correlations. 
}
\label{fig:InfPwrBeta}
\end{figure*}
%%%%%%%%%%%%%%%%%%%%%%%%%%%%%%%%%%%%%%%%%%%%%%%%%%%%%%%%%%%%%%

In Fig. \ref{fig:DL}, we compare DL per-user achievable rates with dynamic TDD and static TDD.
We can see that the DL performance is the worst when static TDD is employed (all cells are in DL).
It is improved as the number of UL cells increases.
In addition, the gap among rates with dynamic TDD and static TDD becomes larger as $M$ increases.
These results support our conclusion that dynamic TDD can achieve better DL performance compared with static TDD.

Fig.~\ref{fig:RateULdiffBeta} shows the UL per-user achievable rates and their deterministic approximations as functions of $M$ under different antenna correlations.
We can see that when $M$ is relatively small, the UL performance of static TDD (all adjacent cells operating in UL) outperforms that of dynamic TDD (some adjacent cells operating in UL while others operating in DL).
The reason is that in this region, the BS-to-BS interference from a DL cell is still more significant than the inter-cell interference from a UL cell.
Thus the more adjacent DL cells there are, the poorer UL performance the considered cell would have.
As $M$ increases, the gaps among rates of dynamic TDD and static TDD are shrinking.
The intersection of the rates in each figure corresponds to the point where dynamic TDD and static TDD achieve the same UL performance.
At those intersections, the BS-to-BS interference from a DL cell is identical to the inter-cell interference from a UL cell.
Thus it makes no difference whether neighboring cells are operating in UL or DL. 
Denote the the number of antennas at the intersection as $M^{*}$.
From the figures, we observe that $M^{*}$ is smaller with lower antenna correlation $\beta$. Specifically, $M^{*}=380,\;240,\;200$ for $\beta=0.4,\;0.2,\;0$, respectively.
In the range of $M>M^{*}$, the redirection of UL transmissions in adjacent cells can reduce the overall interference power to the considered cell,
such that dynamic TDD can improve the UL performance.
When $M=M^{*}$, dynamic TDD is still more beneficial to the network than static TDD since it can offer the flexibility to accommodate different UL/DL traffic patterns across cells.

We now take a closer look at the interference from a single cell. Fig. \ref{fig:InfPwrBeta} compares the post-detection powers of the BS-to-BS
interference from a DL cell and the inter-cell interference from a UL cell under different antenna correlations.
We can observe the following: (i) the powers of the BS-to-BS interference decline dramatically with the increase of $M$,
which validates Proposition 2 and 3,
(ii) When $M$ is relatively small,
the BS-to-BS interference is much more significant than the UL inter-cell interference,
but the gaps between the rates are decreasing rapidly with the growth of $M$, and
(iii) When $M$ is large enough,
the BS-to-BS interference becomes weaker than the UL inter-cell interference, and the gaps grow larger as $M$ increases.
The reason is that in contrast to the BS-to-BS interference which vanishes with infinite $M$, some components of the UL inter-cell interference that represent the pilot contamination effect
would not vanish as $M$ grows \cite{marker5}.
In addition, in each case of antenna correlation, the number of antennas at the point where interference powers intersect perfectly matches $M^{*}$.
This explains the intersections of rates in Fig.~\ref{fig:RateULdiffBeta}.

 %%%%%%%%%%%%%%%%%%%%%%%%%%%%%%%%%%%%%%
\section{Conclusion}
\label{sec:conclusion}
In this work, we proposed a marriage between massive MIMO and dynamic TDD.
Massive MIMO has the potential to address key limitations of dynamic TDD, i.e., the potential increase in interference and loss of UL performance.
We provided detailed analysis based on random matrix theory to show that the effect of the BS-to-BS interference
on uplink transmissions vanishes effectively as the number of BS antennas increases to infinity.
Numerical simulations verified that a dynamic TDD network with massive MIMO achieves higher throughput in both DL and UL compared with a static TDD network.

%%%%%%%%%%%%%%%%%%%%%%%%%%%%%%%%%%%%%%
 \begin{appendices}
\section{Proof of Proposition 3}\label{appendix_A}
\begin{proof}
We will show that both NLoS and LoS components of $I_{jk}^{\mathrm{ul},(4)}$ vanish asymptotically as $M\to \infty$ while $K$ keeps constant. In addition to Assumption 1-4, we assume that $\lim \mathrm{inf}_{M}\;\varphi_{j}^{\mathrm{ul}}\geqslant\delta>0$, which is valid for most MMSE detection designs.

Please refer to Appendix \ref{appendix_B} for useful lemmas used in the following proof.
\subsection{The NLoS components of the BS-to-BS interference}
Each NLoS component of $I_{jk}^{\mathrm{ul},(4)}$ can be written as
\begin{equation}
\mathop{\mathbb{E}} \left[\mathbf{a}_{jk}^{\dagger} \breve{\mathbf{G}}_{jn} \mathbf{W}_{n} \mathbf{W}_{n}^{\dagger} \breve{\mathbf{G}}_{jn}^{\dagger} \mathbf{a}_{jk} \bigg| \hat{\mathbf{H}}_{jj} \right]=\mathbf{a}_{jk}^{\dagger}\mathbf{\Sigma}_{jn}\mathbf{a}_{jk},
\end{equation}where $\mathbf{\Sigma}_{jn}=\mathop{\mathbb{E}}\left[\breve{\mathbf{G}}_{jn} \mathbf{W}_{n} \mathbf{W}_{n}^{\dagger} \breve{\mathbf{G}}_{jn}^{\dagger}\right]$.
It can be bounded as
{\setlength\arraycolsep{1pt}
\begin{eqnarray}
\mathbf{a}_{jk}^{\dagger} \mathbf{\Sigma}_{jn}\mathbf{a}_{jk}&=&\frac{1}{M^{2}}\hat{\mathbf{h}}_{jjk}^{\dagger}\mathbf{\Lambda}_{j}^{\mathrm{ul}} \mathbf{\Sigma}_{jn} \mathbf{\Lambda}_{j}^{\mathrm{ul}} \hat{\mathbf{h}}_{jjk}\nonumber\\
&\stackrel{(a)}{\leqslant}& \left\Vert \mathbf{\Lambda}_{j}^{\mathrm{ul}} \mathbf{\Sigma}_{jn} \mathbf{\Lambda}_{j}^{\mathrm{ul}}\right\Vert \frac{1}{M^{2}}\hat{\mathbf{h}}_{jjk}^{\dagger}\hat{\mathbf{h}}_{jjk}\nonumber\\
&\leqslant&\left\Vert \mathbf{\Sigma}_{jn}\right\Vert \left\Vert \mathbf{\Lambda}_{j}^{\mathrm{ul}}\right\Vert^{2}\frac{1}{M^{2}}\hat{\mathbf{h}}_{jjk}^{\dagger}\hat{\mathbf{h}}_{jjk} \nonumber\\
&\stackrel{(b)}{=}&\frac{\left\Vert \mathbf{\Sigma}_{jn}\right\Vert}{\left\{\min{\mathrm{eig}\left[(\mathbf{\Lambda}_{j}^{\mathrm{ul}})^{-1}\right]}\right\}^{2}}\frac{1}{M^{2}}\hat{\mathbf{h}}_{jjk}^{\dagger}\hat{\mathbf{h}}_{jjk},
\end{eqnarray}}where $ \mathbf{\Lambda}_{j}^{\mathrm{ul}} $ is defined in (\ref{eq:marker1}), (a) follows from Lemma 5, and (b) from Lemma 8. In $(\mathbf{\Lambda}_{j}^{\mathrm{ul}})^{-1}=\frac{1}{M}\sum_{i=1}^{K} \hat{\mathbf{h}}_{jji} \hat{\mathbf{h}}_{jji}^{\dagger} +\frac{1}{M}\mathbf{F}_{j}^{\mathrm{ul}} + \varphi_{j}^{\mathrm{ul}} \mathbf{I}_{M}$, because all terms $\frac{1}{M}\hat{\mathbf{h}}_{jji} \hat{\mathbf{h}}_{jji}^{\dagger}$'s and $\frac{1}{M}\mathbf{F}_{j}^{\mathrm{ul}}$ are positive semi-definite (Assumption 4), their summation is also positive semi-definite and thus all its eigenvalues are nonnegative. From Lemma 9, we have $\mathrm{eig}\left[(\mathbf{\Lambda}_{j}^{\mathrm{ul}})^{-1}\right] \geqslant \varphi_{j}^{\mathrm{ul}} \geqslant\delta$, and therefore
\begin{equation}\label{eq:marker81}
\min{\mathrm{eig}\left[(\mathbf{\Lambda}_{j}^{\mathrm{ul}})^{-1}\right]}\geqslant \delta.
\end{equation}
Then we can obtain,
\begin{equation}
\mathbf{a}_{jk}^{\dagger} \mathbf{\Sigma}_{jn}\mathbf{a}_{jk}\leqslant\frac{1}{\delta^{2}}\left\Vert \mathbf{\Sigma}_{jn}\right\Vert\frac{1}{M^{2}}\hat{\mathbf{h}}_{jjk}^{\dagger}\hat{\mathbf{h}}_{jjk}.
\end{equation}

The next two propositions would show that $\frac{1}{M^{2}}\hat{\mathbf{h}}_{jjk}^{\dagger}\hat{\mathbf{h}}_{jjk}$ converges to zero almost surely, and the spectral norm $\left\Vert \mathbf{\Sigma}_{jn}\right\Vert$ is bounded on $M$, respectively.

\emph{Proposition 4:} Let Assumption 1 hold. Then, $\frac{1}{M^{2}}\hat{\mathbf{h}}_{jjk}^{\dagger}\hat{\mathbf{h}}_{jjk}\xlongrightarrow[M\to \infty]{\mathrm{a.s.}}0$.
\begin{proof}
\begin{eqnarray}\label{eq:marker61}
\frac{1}{M^{2}}\hat{\mathbf{h}}_{jjk}^{\dagger}\hat{\mathbf{h}}_{jjk}&=&\frac{1}{M}\left(\frac{1}{M} \hat{\mathbf{h}}_{jjk}^{\dagger} \hat{\mathbf{h}}_{jjk} - \frac{1}{M} \mathop{\mathrm{tr}} \mathbf{\Phi}_{jjjk} \right) \nonumber\\
&&+ \frac{1}{M^{2}}\mathop{\mathrm{tr}} \mathbf{\Phi}_{jjjk}.
\end{eqnarray}In (\ref{eq:marker61}), from Lemma 4 we have $\frac{1}{M} \hat{\mathbf{h}}_{jjk}^{\dagger} \hat{\mathbf{h}}_{jjk} - \frac{1}{M} \mathop{\mathrm{tr}} \mathbf{\Phi}_{jjjk} \xlongrightarrow[M\to \infty]{\mathrm{a.s.}} 0$. As for the second term,
{\setlength\arraycolsep{1pt}
\begin{eqnarray}\label{eq:marker77}
\frac{1}{M^{2}}\mathop{\mathrm{tr}} \mathbf{\Phi}_{jjjk}
&\stackrel{(a)}{\leqslant}&\frac{1}{M}\left\Vert \mathbf{\Phi}_{jjjk}\right\Vert\nonumber\\
&=& \frac{1}{M}\left\Vert \mathbf{R}_{jjk} \mathbf{Q}_{jk} \mathbf{R}_{jjk}\right\Vert\nonumber\\
&\leqslant&\frac{1}{M}\left\Vert \mathbf{R}_{jjk}\right\Vert^{2} \left\Vert \mathbf{Q}_{jk}\right\Vert \nonumber\\
&\stackrel{(b)}{=}&\frac{1}{M}\frac{\left\Vert \mathbf{R}_{jjk}\right\Vert^{2}}{\min{\mathrm{eig}(\mathbf{Q}_{jk}^{-1})}},
\end{eqnarray}}where $(a)$ follows from Lemma 6, and $(b)$ from Lemma 8. In the expression $\mathbf{Q}_{jk}^{-1}=\sum_{l=1}^{L} \mathbf{R}_{jlk} + \frac{1}{p_{\mathrm{tr}}} \mathbf{I}_{M}$, because each matrix $\mathbf{R}_{jlk}$ is positive semi-definite under Assumption 1, the sum $\sum_{l=1}^{L} \mathbf{R}_{jlk}$ is positive semi-definite as well, and therefore all eigenvalues of $\sum_{l=1}^{L} \mathbf{R}_{jlk}$ are nonnegative. Then from Lemma 9, we have $\mathrm{eig}\left( \mathbf{Q}_{jk}^{-1}\right)=\mathrm{eig}(\sum_{l=1}^{L} \mathbf{R}_{jlk})+\frac{1}{p_{\mathrm{tr}}}\geqslant\frac{1}{p_{\mathrm{tr}}}$, and also $\min{\mathrm{eig}(\mathbf{Q}_{jk}^{-1})}\geqslant\frac{1}{p_{\mathrm{tr}}}$. Then,
\begin{equation}\label{eq:marker78}
\frac{1}{M^{2}}\mathop{\mathrm{tr}} \mathbf{\Phi}_{jjjk}\leqslant \frac{1}{M}\frac{\left\Vert \mathbf{R}_{jjk}\right\Vert^{2}}{\min{\mathrm{eig}(\mathbf{Q}_{jk}^{-1})}} \leqslant \frac{ p_{\mathrm{tr}} R^{2}}{M},
\end{equation}and therefore
\begin{equation}
\frac{1}{M^{2}}\mathop{\mathrm{tr}} \mathbf{\Phi}_{jjjk}\to 0,
\end{equation}as $M\to \infty$, which brings
\begin{equation}\label{eq:marker80}
\frac{1}{M^{2}}\hat{\mathbf{h}}_{jjk}^{\dagger}\hat{\mathbf{h}}_{jjk}\xlongrightarrow[M\to \infty]{\mathrm{a.s.}}0.
\end{equation}
\end{proof}

\emph{Proposition 5:} Let Assumption 1 hold. Then the spectral norm $\left\Vert\mathbf{\Sigma}_{jn}\right\Vert$ is bounded by $KCT$.
\begin{proof}
Denote the $i$th column of $\mathbf{\Omega}_{n}$ by $\boldsymbol{\omega}_{n,i} \in \mathbb{C}^{M}$, where $\mathbf{\Omega}_{n}$ is defined in (\ref{eq:marker5}). Then,
{\setlength\arraycolsep{1pt}
\begin{eqnarray}
\left\Vert\mathbf{\Sigma}_{jn}\right\Vert&=&\left\Vert \mathop{\mathbb{E}}\left[\breve{\mathbf{G}}_{jn} \mathbf{W}_{n} \mathbf{W}_{n}^{\dagger} \breve{\mathbf{G}}_{jn}^{\dagger}\right]\right\Vert\nonumber\\
&\stackrel{(a)}{=}&\left\Vert\bar{\mathbf{C}}_{jn} \mathop{\mathbb{E}} \left[\mathbf{V}_{jn} \bar{\mathbf{T}}_{jn}\mathbf{W}_{n} \mathbf{W}_{n}^{\dagger} \bar{\mathbf{T}}_{jn}^{\dagger} \mathbf{V}_{jn}^{\dagger} \right] \bar{\mathbf{C}}_{jn}^{\dagger} \right\Vert \nonumber\\
&\stackrel{(b)}{=}&\mathop{\mathbb{E}} \left[  \mathop{\mathrm{tr}} \bar{\mathbf{T}}_{jn}\mathbf{W}_{n} \mathbf{W}_{n}^{\dagger}\bar{\mathbf{T}}_{jn}^{\dagger}\right]\left\Vert\mathbf{C}_{jn}\right\Vert\nonumber\\
&=& \lambda_{n} \mathop{\mathbb{E}} \left[  \mathop{\mathrm{tr}} \bar{\mathbf{T}}_{jn}\mathbf{\Omega}_{n} \mathbf{\Omega}_{n}^{\dagger}\bar{\mathbf{T}}_{jn}^{\dagger}\right]\left\Vert\mathbf{C}_{jn}\right\Vert\nonumber\\
&=&\lambda_{n} \mathop{\mathbb{E}} \left[ \sum_{i=1}^{K} \boldsymbol{\omega}_{n,i}^{\dagger} \mathbf{T}_{jn}\boldsymbol{\omega}_{n,i} \right]\left\Vert\mathbf{C}_{jn}\right\Vert,
\end{eqnarray}}where $(a)$ follows from the decomposition (\ref{eq:marker9}), and $(b)$ is obtained by taking the expectation with respect to $\mathbf{V}_{jn}$. By Lemma 5 we can obtain
{\setlength\arraycolsep{1pt}
\begin{eqnarray}
\sum_{i=1}^{K} \boldsymbol{\omega}_{n,i}^{\dagger} \mathbf{T}_{jn}\boldsymbol{\omega}_{n,i} &\leqslant& \Vert \mathbf{T}_{jn}\Vert \sum_{i=1}^{K} \Vert\boldsymbol{\omega}_{n,i} \Vert_{2}^{2} \nonumber\\
&=&\Vert \mathbf{T}_{jn}\Vert \mathop{\mathrm{tr}} \mathbf{\Omega}_{n} \mathbf{\Omega}_{n}^{\dagger},
\end{eqnarray}}which holds with the mean as well, i.e.,
\begin{equation}
\mathop{\mathbb{E}} \left[ \sum_{i=1}^{K} \boldsymbol{\omega}_{n,i}^{\dagger} \mathbf{T}_{jn}\boldsymbol{\omega}_{n,i} \right] \leqslant \Vert \mathbf{T}_{jn}\Vert \mathop{\mathbb{E}} \left[\mathop{\mathrm{tr}} \mathbf{\Omega}_{n} \mathbf{\Omega}_{n}^{\dagger} \right].
\end{equation}Then,
{\setlength\arraycolsep{1pt}
\begin{eqnarray}
\left\Vert \mathbf{\Sigma}_{jn}\right\Vert &\leqslant& \lambda_{n}\mathop{\mathbb{E}} \left[\mathop{\mathrm{tr}} \mathbf{\Omega}_{n} \mathbf{\Omega}_{n}^{\dagger} \right] \Vert \mathbf{T}_{jn}\Vert \Vert\mathbf{C}_{jn}\Vert \nonumber\\
&\stackrel{(a)}{=}&K \Vert \mathbf{T}_{jn}\Vert \Vert\mathbf{C}_{jn}\Vert \nonumber\\
&\leqslant&KCT,
\end{eqnarray}}where $(a)$ follows from the definition $\lambda_{n}=K/\mathop{\mathbb{E}} \left[ \mathop{\mathrm{tr}} \mathbf{\Omega}_{n} \mathbf{\Omega}_{n}^{\dagger} \right]$.
\end{proof}

Combing Proposition 4 and 5, we can obtain
\begin{equation}
\mathbf{a}_{jk}^{\dagger} \mathbf{\Sigma}_{jn}\mathbf{a}_{jk}\leqslant\frac{KCT}{\delta^{2}}\frac{1}{M^{2}}\hat{\mathbf{h}}_{jjk}^{\dagger}\hat{\mathbf{h}}_{jjk},
\end{equation}and therefore
\begin{equation}
\mathbf{a}_{jk}^{\dagger} \mathbf{\Sigma}_{jn}\mathbf{a}_{jk}\xlongrightarrow[M\to \infty]{\mathrm{a.s.}} 0,
\end{equation}i.e., each NLoS component of the BS-to-BS interference converges to zero almost surely as $M\to \infty$.

\subsection{The LoS components of the BS-to-BS interference}
The LoS components of $I_{jk}^{\mathrm{ul},(4)}$ will be analyzed as follows. Define
\begin{equation}\label{eq:marker65}
\mathbf{\Lambda}_{jk}^{\mathrm{ul}}=\left( \frac{1}{M}\sum_{\scriptstyle i=1 \atop \scriptstyle i\neq k}^{K} \hat{\mathbf{h}}_{jji} \hat{\mathbf{h}}_{jji}^{\dagger} +\frac{1}{M}\mathbf{F}_{j}^{\mathrm{ul}} + \varphi_{j}^{\mathrm{ul}} \mathbf{I}_{M} \right)^{-1}.
\end{equation}By using similar approaches as in the proof of (\ref{eq:marker81}), we can obtain
\begin{equation}\label{eq:marker79}
\min{\mathrm{eig}\left[(\mathbf{\Lambda}_{jk}^{\mathrm{ul}})^{-1}\right]}\geqslant\delta >0.
\end{equation}Then all eigenvalues of $\mathbf{\Lambda}_{jk}^{\mathrm{ul}}$ are positive, and therefore $\mathbf{\Lambda}_{jk}^{\mathrm{ul}}$ is positive definite. Each LoS component of $I_{jk}^{\mathrm{ul},(4)}$ takes the form
{\setlength\arraycolsep{1pt}
\begin{eqnarray}
&&\mathop{\mathbb{E}} \left[\mathbf{a}_{jk}^{\dagger} \bar{\mathbf{G}}_{jn} \mathbf{W}_{n} \mathbf{W}_{n}^{\dagger} \bar{\mathbf{G}}_{jn}^{\dagger} \mathbf{a}_{jk} \bigg| \hat{\mathbf{H}}_{jj} \right]\nonumber\\
&=&\frac{1}{M^{2}}\hat{\mathbf{h}}_{jjk}^{\dagger} \mathbf{\Lambda}_{j}^{\mathrm{ul}} \bar{\mathbf{G}}_{jn}\mathop{\mathbb{E}} \left[\mathbf{W}_{n}\mathbf{W}_{n}^{\dagger} \right]\bar{\mathbf{G}}_{jn}^{\dagger}\mathbf{\Lambda}_{j}^{\mathrm{ul}} \hat{\mathbf{h}}_{jjk}\nonumber\\
&\stackrel{(a)}{=}&\frac{1}{M^{2}}\frac{\hat{\mathbf{h}}_{jjk}^{\dagger} \mathbf{\Lambda}_{jk}^{\mathrm{ul}} \bar{\mathbf{G}}_{jn}\mathop{\mathbb{E}} \left[\mathbf{W}_{n}\mathbf{W}_{n}^{\dagger} \right]\bar{\mathbf{G}}_{jn}^{\dagger}\mathbf{\Lambda}_{jk}^{\mathrm{ul}} \hat{\mathbf{h}}_{jjk}}{\left(1+\frac{1}{M}\hat{\mathbf{h}}_{jjk}^{\dagger} \mathbf{\Lambda}_{jk}^{\mathrm{ul}}\hat{\mathbf{h}}_{jjk}\right)^{2}} \nonumber\\
&\stackrel{(b)}{\leqslant}& \frac{1}{M^{2}}\hat{\mathbf{h}}_{jjk}^{\dagger} \mathbf{\Lambda}_{jk}^{\mathrm{ul}} \bar{\mathbf{G}}_{jn}\mathop{\mathbb{E}} \left[\mathbf{W}_{n}\mathbf{W}_{n}^{\dagger} \right]\bar{\mathbf{G}}_{jn}^{\dagger}\mathbf{\Lambda}_{jk}^{\mathrm{ul}} \hat{\mathbf{h}}_{jjk} \nonumber\\
&=&\frac{1}{M}\Big( \frac{1}{M}\hat{\mathbf{h}}_{jjk}^{\dagger} \mathbf{\Lambda}_{jk}^{\mathrm{ul}} \bar{\mathbf{G}}_{jn}\mathop{\mathbb{E}} \left[\mathbf{W}_{n}\mathbf{W}_{n}^{\dagger} \right]\bar{\mathbf{G}}_{jn}^{\dagger}\mathbf{\Lambda}_{jk}^{\mathrm{ul}} \hat{\mathbf{h}}_{jjk} \nonumber\\
&&-\frac{1}{M}\mathop{\mathrm{tr}}\mathbf{\Phi}_{jjjk}\mathbf{\Lambda}_{jk}^{\mathrm{ul}}\bar{\mathbf{G}}_{jn}\mathop{\mathbb{E}} \left[\mathbf{W}_{n}\mathbf{W}_{n}^{\dagger} \right]\bar{\mathbf{G}}_{jn}^{\dagger}\mathbf{\Lambda}_{jk}^{\mathrm{ul}} \Big) \nonumber\\
&&+\frac{1}{M^{2}}\mathop{\mathrm{tr}}\mathbf{\Phi}_{jjjk}\mathbf{\Lambda}_{jk}^{\mathrm{ul}}\bar{\mathbf{G}}_{jn}\mathop{\mathbb{E}} \left[\mathbf{W}_{n}\mathbf{W}_{n}^{\dagger} \right]\bar{\mathbf{G}}_{jn}^{\dagger}\mathbf{\Lambda}_{jk}^{\mathrm{ul}},
\end{eqnarray}}where $(a)$ follows from Lemma 3, and $(b)$ holds true because $\mathbf{\Lambda}_{jk}^{\mathrm{ul}}$ is positive definite such that $\frac{1}{M}\hat{\mathbf{h}}_{jjk}^{\dagger} \mathbf{\Lambda}_{jk}^{\mathrm{ul}}\hat{\mathbf{h}}_{jjk}\geqslant 0$ and $\left(1+\frac{1}{M}\hat{\mathbf{h}}_{jjk}^{\dagger} \mathbf{\Lambda}_{jk}^{\mathrm{ul}}\hat{\mathbf{h}}_{jjk}\right)^{2}\geqslant 1$. In the last equation, by Lemma 4, the first term converges to zero almost surely. For the second term,
{\setlength\arraycolsep{1pt}
\begin{eqnarray}
&&\frac{1}{M^{2}}\mathop{\mathrm{tr}}\mathbf{\Phi}_{jjjk}\mathbf{\Lambda}_{jk}^{\mathrm{ul}}\bar{\mathbf{G}}_{jn}\mathop{\mathbb{E}} \left[\mathbf{W}_{n}\mathbf{W}_{n}^{\dagger} \right]\bar{\mathbf{G}}_{jn}^{\dagger}\mathbf{\Lambda}_{jk}^{\mathrm{ul}}\nonumber\\
&=&\frac{1}{M^{2}}\mathop{\mathrm{tr}}\bar{\mathbf{G}}_{jn}^{\dagger}\mathbf{\Lambda}_{jk}^{\mathrm{ul}}\mathbf{\Phi}_{jjjk}\mathbf{\Lambda}_{jk}^{\mathrm{ul}}\bar{\mathbf{G}}_{jn}\mathop{\mathbb{E}} \left[\mathbf{W}_{n}\mathbf{W}_{n}^{\dagger} \right]\nonumber\\
&\stackrel{(a)}{\leqslant}&\frac{1}{M^{2}}\left\Vert \bar{\mathbf{G}}_{jn}^{\dagger}\mathbf{\Lambda}_{jk}^{\mathrm{ul}}\mathbf{\Phi}_{jjjk}\mathbf{\Lambda}_{jk}^{\mathrm{ul}}\bar{\mathbf{G}}_{jn}\right\Vert \mathop{\mathrm{tr}}\mathop{\mathbb{E}} \left[\mathbf{W}_{n}\mathbf{W}_{n}^{\dagger} \right] \nonumber\\
&=&\frac{1}{M^{2}}\left\Vert \bar{\mathbf{G}}_{jn}^{\dagger}\mathbf{\Lambda}_{jk}^{\mathrm{ul}}\mathbf{\Phi}_{jjjk}\mathbf{\Lambda}_{jk}^{\mathrm{ul}}\bar{\mathbf{G}}_{jn}\right\Vert
\lambda_{n}\mathop{\mathbb{E}}\left[\mathop{\mathrm{tr}} \mathbf{\Omega}_{n} \mathbf{\Omega}_{n}^{\dagger}\right]\nonumber\\
&\stackrel{(b)}{=}&\frac{K}{M^{2}}\left\Vert \bar{\mathbf{G}}_{jn}^{\dagger}\mathbf{\Lambda}_{jk}^{\mathrm{ul}}\mathbf{\Phi}_{jjjk}\mathbf{\Lambda}_{jk}^{\mathrm{ul}}\bar{\mathbf{G}}_{jn}\right\Vert
\nonumber\\
&\leqslant&\frac{K}{M^{2}}\left\Vert \mathbf{\Lambda}_{jk}^{\mathrm{ul}}\right\Vert^{2} \left\Vert \bar{\mathbf{G}}_{jn}^{\dagger} \bar{\mathbf{G}}_{jn}\right\Vert \left\Vert \mathbf{\Phi}_{jjjk}\right\Vert \nonumber\\
&\stackrel{(c)}{=}&\frac{K}{M^{2}}\frac{1}{\left\{\min{\mathrm{eig}\left[(\mathbf{\Lambda}_{jk}^{\mathrm{ul}})^{-1}\right]}\right\}^{2}} \left\Vert \bar{\mathbf{G}}_{jn}^{\dagger} \bar{\mathbf{G}}_{jn}\right\Vert \left\Vert \mathbf{\Phi}_{jjjk}\right\Vert\nonumber\\
&\stackrel{(d)}{\leqslant}&\frac{p_{\mathrm{tr}} GR^{2}K}{\delta^{2}M},
\end{eqnarray}}where $(a)$ follows from Lemma 7, $(b)$ follows from the definition of $\lambda_{n}$, $(c)$ follows from Lemma 8, and $(d)$ follows from Assumption 3, (\ref{eq:marker79}), (\ref{eq:marker77}) and (\ref{eq:marker78}). Then,
\begin{equation}
\frac{1}{M^{2}}\mathop{\mathrm{tr}}\mathbf{\Phi}_{jjjk}\mathbf{\Lambda}_{jk}^{\mathrm{ul}}\bar{\mathbf{G}}_{jn}\mathop{\mathbb{E}} \left[\mathbf{W}_{n}\mathbf{W}_{n}^{\dagger} \right]\bar{\mathbf{G}}_{jn}^{\dagger}\mathbf{\Lambda}_{jk}^{\mathrm{ul}}\to 0
\end{equation}as $M \to \infty$, and thus we have
\begin{equation}\label{eq:marker68}
\mathop{\mathbb{E}} \left[\mathbf{a}_{jk}^{\dagger} \bar{\mathbf{G}}_{jn} \mathbf{W}_{n} \mathbf{W}_{n}^{\dagger} \bar{\mathbf{G}}_{jn}^{\dagger} \mathbf{a}_{jk} \bigg| \hat{\mathbf{H}}_{jj} \right]\xlongrightarrow[M\to \infty]{\mathrm{a.s.}} 0,
\end{equation}i.e., each LoS component of the BS-to-BS interference converges to zero almost surely as $M\to \infty$.
\end{proof}
%%%%%%%%%%%%%%%%%%%%%%%%%%%%%%%%%%%%%%%
\section{Useful Lemmas}\label{appendix_B}
\emph{Lemma 1 (Theorem 1 \cite{marker6}):} Assume that $\mathbf{D} \in \mathbb{C}^{N \times N}$ and $\mathbf{S} \in \mathbb{C}^{N \times N}$ are Hermitian nonnegative definite matrices and $\mathbf{h}_{i} \in \mathbb{C}^{N}, i=1,\ldots,n$ are random vectors subject to $\mathcal{CN}\left( \mathbf{0},\frac{1}{N}\mathbf{R}_{i} \right)$. 
Suppose that spectral norms of $\mathbf{D}$ and $\mathbf{R}_{i}$'s are uniformly bounded with respect to $N$. Then for any $\rho>0$, we have
\begin{equation}
\frac{1}{N} \mathop{\mathrm{tr}}\mathbf{D}\left( \sum_{i=1}^{n} \mathbf{h}_{i}\mathbf{h}_{i}^{\dagger} + \mathbf{S} +\rho \mathbf{I}_{N} \right)^{-1} - \frac{1}{N}\mathop{\mathrm{tr}}\mathbf{D} \mathbf{\Gamma}(\rho) \xlongrightarrow[N\to \infty]{\mathrm{a.s.}} 0, \nonumber
\end{equation}where 
\begin{equation}
\mathbf{\Gamma}(\rho)=\left( \frac{1}{N}\sum_{i=1}^{n} \frac{\mathbf{R}_{i}}{1+\delta_{i}(\rho)}+\mathbf{S}+\rho \mathbf{I}_{N} \right)^{-1}, \nonumber
\end{equation}and $\delta_{i}(\rho)$, $i=1,\ldots,n$ are defined as $\delta_{i}(\rho)=\lim_{t \rightarrow \infty} \delta_{i}^{(t)}(\rho)$, where for $t=1,2,\ldots,$
\begin{equation}
\delta_{i}^{(t)}(\rho)=\frac{1}{N}\mathop{\mathrm{tr}} \mathbf{R}_{i} \left( \frac{1}{N}\sum_{m=1}^{n} \frac{\mathbf{R}_{m}}{1+\delta_{m}^{(t-1)}(\rho)}+\mathbf{S}+\rho \mathbf{I}_{N} \right)^{-1}, \nonumber
\end{equation}with initial values $\delta_{i}^{(0)}(\rho)=1/\rho, \forall i$. $\delta_{i}(\rho)$'s can be calculated efficiently by the fixed-point algorithm with guaranteed convergence.

\emph{Lemma 2 (Theorem 2 \cite{marker6}):} Let $\mathbf{\Theta} \in \mathbb{C}^{N \times N}$ be Hermitian nonnegative definite with uniformly bounded spectral norm with respect to $N$. Under the conditions of Lemma 1, we have
{\setlength\arraycolsep{1pt}
\begin{eqnarray}
\frac{1}{N} \mathop{\mathrm{tr}}\mathbf{D}( \sum_{i=1}^{n} \mathbf{h}_{i}\mathbf{h}_{i}^{\dagger} + \mathbf{S} +\rho \mathbf{I}_{N} )^{-1} \mathbf{\Theta} ( \sum_{i=1}^{n} \mathbf{h}_{i}\mathbf{h}_{i}^{\dagger} + \mathbf{S} +\rho \mathbf{I}_{N} )^{-1} &&\nonumber\\
- \frac{1}{N}\mathop{\mathrm{tr}}\mathbf{D} \mathbf{\Gamma}^{\prime}(\rho) \xlongrightarrow[N\to \infty]{\mathrm{a.s.}} 0, &&\nonumber
\end{eqnarray}}where 
\begin{equation}
\mathbf{\Gamma}^{\prime}(\rho)=\mathbf{\Gamma}(\rho) \mathbf{\Theta} \mathbf{\Gamma}(\rho) + \mathbf{\Gamma}(\rho) \left[ \frac{1}{N} \sum_{i=1}^{n} \frac{\mathbf{R}_{i} \delta_{i}^{\prime}(\rho)}{(1+\delta_{i}(\rho))^{2}} \right] \mathbf{\Gamma}(\rho), \nonumber
\end{equation}
$\mathbf{\Gamma}(\rho)$ and $\delta_{i}(\rho)$ are given by Lemma 1, and $\boldsymbol{\delta}^{\prime}(\rho)=[\delta_{1}^{\prime}(\rho) \ldots \delta_{n}^{\prime}(\rho)]^{\boldsymbol{\top}}$ which is given by
\begin{equation}
\boldsymbol{\delta}^{\prime}(\rho) = (\mathbf{I}_{n}-\mathbf{J}(\rho))^{-1} \mathbf{u}(\rho), \nonumber
\end{equation}
where
\begin{equation}
[\mathbf{J}(\rho)]_{r,c} = \frac{ \frac{1}{N} \mathop{\mathrm{tr}} \mathbf{R}_{r} \mathbf{\Gamma}(\rho) \mathbf{R}_{c} \mathbf{\Gamma}(\rho)}{N(1+\delta_{c}(\rho))^{2}} \quad 1 \leq r,c \leq n, \nonumber
\end{equation}
\begin{equation}
[\mathbf{u}(\rho)]_{r} = \frac{1}{N} \mathop{\mathrm{tr}} \mathbf{R}_{r} \mathbf{\Gamma}(\rho) \mathbf{\Theta} \mathbf{\Gamma}(\rho) \quad 1 \leq r \leq n. \nonumber
\end{equation}

\emph{Lemma 3 (Matrix Inversion Lemma \cite{marker6, marker9}):} Let $\mathbf{A} \in \mathbb{C}^{N \times N}$ be Hermitian invertible. Then, for any vector $\mathbf{x} \in \mathbb{C}^{N}$ and any scalar $\tau \in \mathbb{C}$ such that $\mathbf{A} + \tau \mathbf{x} \mathbf{x}^{\dagger}$ is invertible,
\begin{equation}
\mathbf{x}^{\dagger} \left( \mathbf{A}+\tau \mathbf{x} \mathbf{x}^{\dagger} \right)^{-1} = \frac{\mathbf{x}^{\dagger}\mathbf{A}^{-1}}{1+\tau \mathbf{x}^{\dagger}\mathbf{A}^{-1}\mathbf{x}}.\nonumber
\end{equation}

\emph{Lemma 4 (\cite{marker6, marker9}):} Let $\mathbf{A} \in \mathbb{C}^{N \times N}$ and $\mathbf{x}, \mathbf{y} \sim \mathcal{CN}\left( \mathbf{0},\frac{1}{N}\mathbf{I}_{N} \right)$. Assume that $\mathbf{A}$ has uniformly bounded spectral norm (with respect to $N$) and that $\mathbf{x}$ and $\mathbf{y}$ are mutually independent and independent of $\mathbf{A}$. Then we have
{\setlength\arraycolsep{1pt}
\begin{eqnarray}
\mathbf{x}^{\dagger} \mathbf{A} \mathbf{x} - \frac{1}{N} \mathop{\mathrm{tr}} \mathbf{A} & \xlongrightarrow[N\to \infty]{\mathrm{a.s.}} & 0, \nonumber\\
\mathbf{x}^{\dagger} \mathbf{A} \mathbf{y} & \xlongrightarrow[N\to \infty]{\mathrm{a.s.}} & 0.\nonumber
\end{eqnarray}}

\emph{Lemma 5:} Let $\mathbf{A} \in \mathbb{C}^{N \times N}$ be Hermitian nonnegative definite and $\mathbf{x} \in \mathbb{C}^{N}$. Then
\begin{equation}
\mathbf{x}^{\dagger} \mathbf{A} \mathbf{x}  \leqslant  \Vert \mathbf{A} \Vert \Vert \mathbf{x} \Vert _{2}^{2}. \nonumber
\end{equation}

\emph{Lemma 6:} Let $\mathbf{A} \in \mathbb{C}^{N \times N}$ be Hermitian nonnegative definite. Then
\begin{equation}
\frac{1}{N} \mathop{\mathrm{tr}} \mathbf{A}  \leqslant  \Vert \mathbf{A} \Vert. \nonumber
\end{equation}

\emph{Lemma 7:} Let the matrices $\mathbf{A}$ and $\mathbf{B}$ be Hermitian nonnegative definite. Then
\begin{equation}
\mathop{\mathrm{tr}}\mathbf{A}\mathbf{B}\leqslant \Vert \mathbf{A}\Vert \mathop{\mathrm{tr}}\mathbf{B}. \nonumber
\end{equation}

\emph{Lemma 8:} Let the matrix $\mathbf{A}$ be Hermitian invertible,
\begin{equation}
\Vert \mathbf{A}^{-1} \Vert = \max{\mathrm{eig}(\mathbf{A}^{-1})}=\frac{1}{\min{\mathrm{eig}(\mathbf{A})}}. \nonumber
\end{equation}

\emph{Lemma 9:} Let $\mathbf{A}\in \mathbb{C}^{N \times N}$ be Hermitian and $c \in \mathbb{R}$. Then,
\begin{equation}
\mathrm{eig}(\mathbf{A}+c\mathbf{I}_{N}) = \mathrm{eig}(\mathbf{A})+c. \nonumber
\end{equation}
\end{appendices}

\end{document}